\documentclass[conference]{IEEEtran}
\IEEEoverridecommandlockouts
\usepackage{cite}
\usepackage{amsmath,amssymb,amsfonts}
\usepackage{algorithmic}
\usepackage{graphicx}
\usepackage{textcomp}
\usepackage{xcolor}
\usepackage{threeparttable}
\usepackage{url}
\usepackage{siunitx}
\def\BibTeX{{\rm B\kern-.05em{\sc i\kern-.025em b}\kern-.08em
    T\kern-.1667em\lower.7ex\hbox{E}\kern-.125emX}}

\usepackage{fancyhdr}
\fancypagestyle{mahmood}{%
  \fancyhf{} 
  
  \fancyhead[C]{\footnotesize \textcopyright 2022 IEEE. Personal use of this material is permitted.  Permission from IEEE must be obtained for all other uses, in any current or future media, including reprinting/republishing this material for advertising or promotional purposes, creating new collective works, for resale or redistribution to servers or lists, or reuse of any copyrighted component of this work in other works.}
}%
\makeatletter
\let\ps@IEEEtitlepagestyle\ps@mahmood
\makeatother
\begin{document}

\title{Exploring Automatic Gym Workouts Recognition Locally On Wearable Resource-Constrained Devices\\
}

\author{\IEEEauthorblockN{ Sizhen Bian}
\IEEEauthorblockA{\textit{PBL D-ITET} \\
\textit{ETH Zürich}\\
Zürich, Switzerland \\
sizhen.bian@pbl.ee.ethz.ch}
\and
\IEEEauthorblockN{ Xiaying Wang}
\IEEEauthorblockA{\textit{PBL D-ITET} \\
\textit{ETH Zürich}\\
Zürich, Switzerland \\
xiaywang@iis.ee.ethz.ch}
\and
\IEEEauthorblockN{Tommaso Polonelli}
\IEEEauthorblockA{\textit{PBL D-ITET} \\
\textit{ETH Zürich}\\
Zürich, Switzerland \\
tommaso.polonelli@pbl.ee.ethz.ch}
\and
\IEEEauthorblockN{ Michele Magno}
\IEEEauthorblockA{\textit{PBL D-ITET} \\
\textit{ETH Zürich}\\
Zürich, Switzerland \\
michele.magno@pbl.ee.ethz.ch}
}

\IEEEoverridecommandlockouts
\IEEEpubid{\makebox[\columnwidth]{978-1-6654-6550-2/22/\$31.00~\copyright2022 IEEE \hfill} \hspace{\columnsep}\makebox[\columnwidth]{ }}

\maketitle

\IEEEpubidadjcol

\begin{abstract}
Automatic gym activity recognition on energy- and resource-constrained wearable devices removes the human-interaction requirement during intense gym sessions - like soft-touch tapping and swiping. This work presents a tiny and highly accurate residual convolutional neural network that runs in milliwatt microcontrollers for automatic workouts classification. We evaluated the inference performance of the deep model with quantization on three resource-constrained devices: two microcontrollers with ARM-Cortex M4 and M7 core from ST Microelectronics, and a GAP8 system on chip, which is an open-sourced, multi-core RISC-V computing platform from GreenWaves Technologies. Experimental results show an accuracy of up to 90.4\% for eleven workouts recognition with full precision inference. The paper also presents the trade-off performance of the resource-constrained system. While keeping the recognition accuracy (88.1\%) with minimal loss, each inference takes only \SI{3.2}ms on GAP8, benefiting from the 8 RISC-V cluster cores. We measured that it features an execution time that is 18.9x and 6.5x faster than the Cortex-M4 and Cortex-M7 cores, showing the feasibility of real-time on-board workouts recognition based on the described data set with \SI{20} Hz sampling rate. The energy consumed for each inference on GAP8 is \SI{0.41}mJ compared to \SI{5.17}mJ on Cortex-M4 and \SI{8.07}mJ on Cortex-M7 with the maximum clock. It can lead to longer battery life when the system is battery-operated. We also introduced an open data set composed of fifty sessions of eleven gym workouts collected from ten subjects that is publicly available. 

\end{abstract}


\section{Introduction}

Current commercial smart wearable devices are capable of helping people to track their fitness activities. On the other hand, they usually enable only a limited number of fitness types that can be recognized automatically without the user's input~\cite{khurana2019past}. For example, the Fitbit smart wristband only identifies and records a few activities, like running and swimming. Moreover, it needs at least fifteen minutes to trigger the automatic recognition feature \cite{leung2022meta}. A popular commercial solution for recording the fitness activity is the fitness app~\cite{radhakrishnan2020gym}, which requires user interventions like swiping and tipping on the wearable screen. These limitations are mainly due to the lack of wearable computing ability w.r.t. the complexity of body actions \cite{wang2020fann}. Since current dominant wearable motion sensing is solely based on the inertial measurement unit (IMU) \cite{ishii2021exersense}, a typical wrist-worn smart device only captures the motion pattern of the wrist. Achieving a more smart device with onboard accurate inference capability, long lifetime, and low latency is still a challenging task, and it requires a good trade-off between energy efficiency and computational performance \cite{wang2020fann}. 
 
Automatic gym activity tracking provides the user with an intense gym experience and a reliable historical workout overview, thus improving the overall well-being of a person. A wrist-worn device could hardly sense the leg-dominated workouts, like the adductor machine, unless multiple IMUs are deployed on the limbs~\cite{chang2007tracking}. The non-wearable sensing modalities, like pressure mat~\cite{sundholm2014activity}, radiofrequency signal~\cite{guo2018device}, and vision-based ones~\cite{ganesh2020personalized,vyas2019pose}, could supply much higher recognition accuracy or unlimited recognized workout types but result in user inconveniences like an extra burden or privacy issues. Another reason for wearable devices' limited workout types recognition is the limitation of local computing ability \cite{wang2020fann}. Since the success of deep neural networks in computer vision and natural language processing, researchers have explored the performance of this featuring technology on sensor-based data (which is usually one-dimensional data stream or multi-channel one-dimensional data streams from fused sensors) for human activity recognition, and got impressive result in tasks like facial expression recognition\cite{revina2021survey}, daily living activities recognition\cite{hussain2019performance}, hand gesture recognition\cite{jaramillo2020real}, etc. However, since the inference of deep models needs millions of multiply-accumulate operations and millions of weight and activation parameters, a wearable device usually is too weak to support the inferences regarding its memory space, clock, and computing ability. 

Considering the massive deployment of Internet of Things (IoT) devices around human life, researchers have focused on skills aiming to run a neural network architecture on those resource-limited devices with kilobytes of RAM by compressing an exited model with minimal accuracy loss. For example, quantization~\cite{zhu2016trained}, changing the model parameters from floating-point numbers to low-bit width numbers; Pruning\cite{han2015deep}, removing the least important parameters that do not affect the decision-making process of the model; Knowledge distillation~\cite{bucilu2006model}, transferring the knowledge to the student model form a teacher model. Those model compression approaches can be considered individually as well as jointly~\cite{gil2021quantization,choi2020data} and have shown their validity in enormous use scenarios~\cite{bian2021capacitive, ingolfsson2021ecg, wang2020accurate}. Besides that, novel hardware architectures were also developed to bring intelligence to the edge. Such hardware platforms consume milliwatt of power and are suitable for wearable battery-operated devices\cite{louis2019towards, garofalo2020pulp, sorbaro2020optimizing}. Among them, the novel GAP8 architecture, which features 8 RISC-V cores, has shown impressive inference performance in certain human activity recognition tasks\cite{benninger2019edgeeye, wang2020fann}. This triggered the idea of automatic gym workout recognition and recording locally on cheap and ultra resource-constrained devices, aiming to enable a better user experience for gym enthusiasts.

This paper explored a lightweight residual tiny neural network model for gym workouts recognition on energy efficient embedded devices and demonstrated its work efficiency on three resource-constrained platforms. Following contributions are presented:

\begin{enumerate}
\item We first introduced a gym data set including eleven workouts for classification, collected from ten volunteers in multiple days with sensors deployed at the wrist, leg, and pocket. The data set is public available\footnote{\url{https://github.com/zhaxidele/Toolkit-for-HBC-sensing}} aiming to promote other researchers for further exploration. The sensing modalities of the data set include the traditional inertial measurement sensors and a novel human body capacitance signal~\cite{bian2021systematic}, which has been demonstrated with impressive alternative or complementary role for body motion sensing~\cite{bian2019passive}.  
\item We utilized the residual neural network model and deployed the compressed model on three processors that are designed for embedded applications: the ARM Cortex-M4/M7 MCU from STMicroelectronics and the RISC-V PULP-based GAP8 MCU\cite{flamand2018gap} from GreenWaves Technologies. All three platforms show minimal accuracy drop for the inferences with 8-bits weights and activation parameters.
\item We demonstrated the possibility to enable automatic activity recognition directly on-board by exploring the inference performance on the selected platforms regarding the throughput, latency, and power efficiency. While setting all three platforms with their maximum clock speed, the PULP platform takes only \SI{3.2}ms for each inference, which is 18.9x and 6.5x faster than the Cortex-M4 and Cortex-M7 cores, showing the feasibility of real-time on-board workouts recognition on the described data set with 20Hz sensor data sampling rate. The multiply and accumulation (MAC)  operations in each cycle also show a 13.0x and 8.1x improvement on GAP8. 
\item We demonstrated the advantage of the proposed algorithm on the selected GAP8 platform in terms of energy efficiency. Experimental measurements show the energy consumed for each inference on GAP8 is \SI{0.41}mJ while \SI{5.17}mJ on Cortex-M4 and \SI{8.07}mJ on Cortex-M7. 
\end{enumerate}

\begin{figure}[t]
\centering
\includegraphics[width=0.95\linewidth,height=8cm]{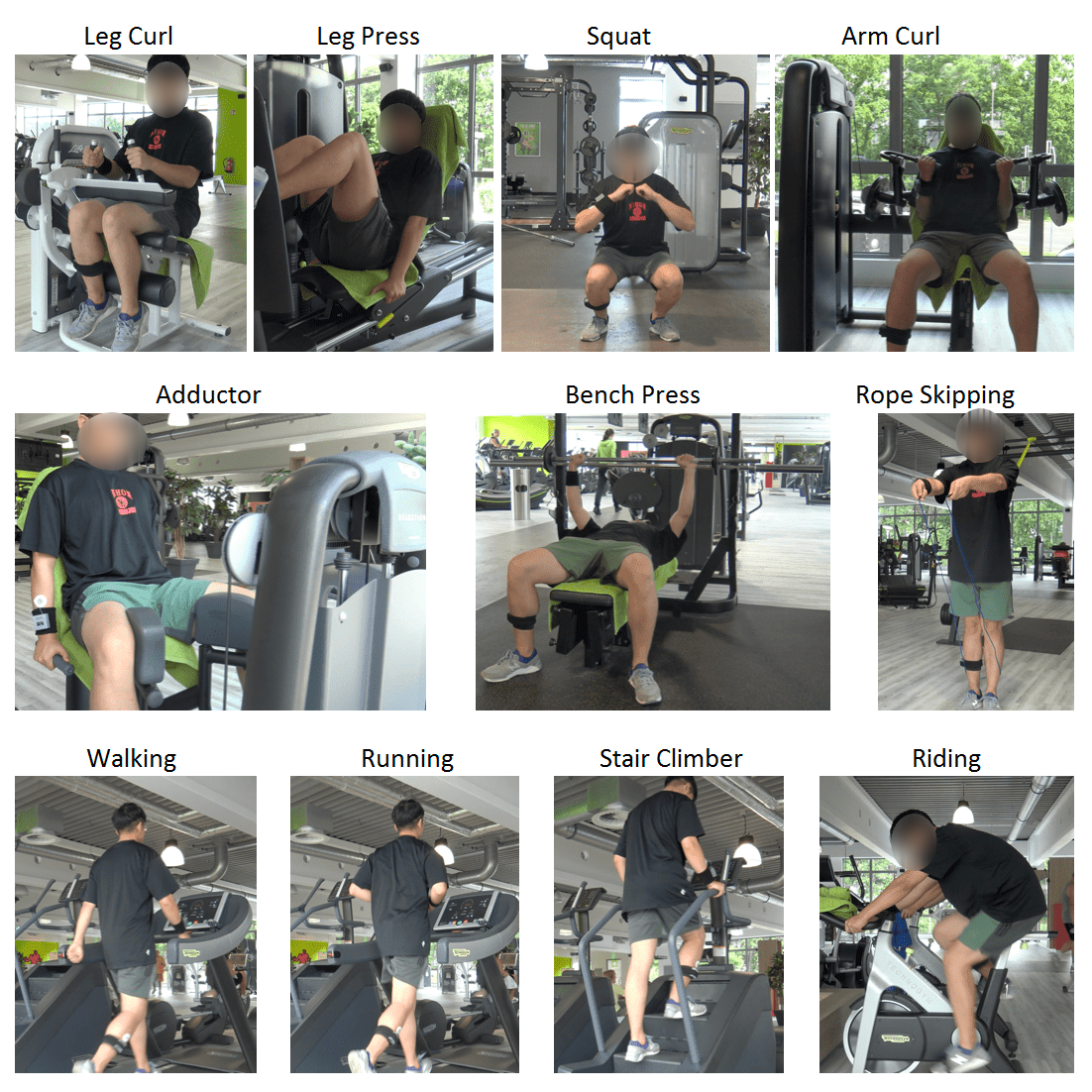}
\caption{Eleven gym workouts}
\label{Eleven}
\end{figure}

\section{Data Set and Set-up}

The presented data set includes eleven popular aerobic and anaerobic workouts that are supported in a typical gym studio: Adductor, Armcurl, Benchpress, Legcurl, Legpress, Riding, Ropeskipping, Running, Squat, Stairsclimber, and Walking (as Fig. \ref{Eleven} depicts). Running and Walking were performed on the treadmills with the speed of 5$\pm$0.2 km/h and 8$\pm$0.5 km/h for around 2 minutes in each session. Riding and Stairclimber were done at a self-determined pace and lasted about 2 minutes. The rest exercises were trained with gym instruments (except Squat) for 3x10 repetitions. Ten volunteers participated in this study, including five females and five males. Eight of them go to the gym at least three times a week, and two of them are novices. Each participant performed the above-listed exercises for five sessions in five days.

\begin{figure*}[t]
\centering
\includegraphics[width=0.8\textwidth,height=10.5cm]{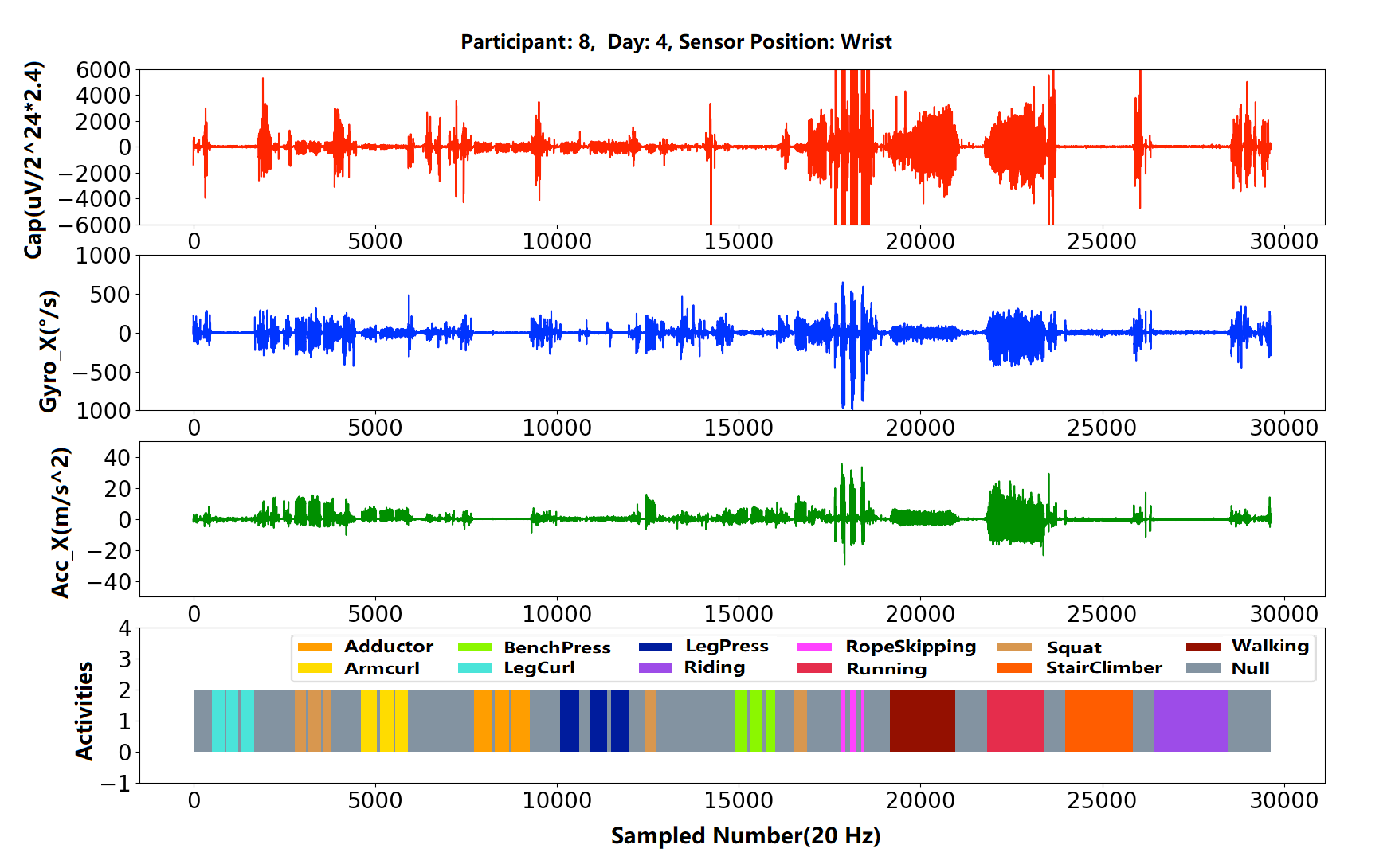}
\caption{Example of one session's initial measurement unit signal, body capacitance caused potential variation signal and the exercise labels, including a null class.}
\label{Session}
\end{figure*}

The sensing unit contains both inertial measurement unit (IMU) and the human body capacitance (HBC) sensing unit described in \cite{bian2022using}. Human body capacitance is a particular property describing the capacitance between the body and the ground or environment, a leg motion causes an overall body capacitance variation, and the variation could be captured by the body capacitance sensing unit worn on the wrist\cite{bian2019wrist}. The two sensing units were integrated on a small PCB powered by a 3.7V lithium battery for the gym experiment data collection. The data set includes data collected with three sensor positions: the wrist, the leg, and the pocket. Since this paper only focuses on exploring local inference of the workouts, we only use the data with the sensor worn on the wrist for the purpose. 

Fig. \ref{Session} depicts one whole session of IMU and HBC data from the eighth volunteer in the fourth training day with the prototype worn on the wrist. The grey color in the fourth subplot is defined as "Null" state, which is also being classified by the neural network model. Data was sampled with a frequency of 20 Hz. Overall, fifty sessions data from the ten volunteers were collected.

\section{Proposed TinyML Residual Model}
\begin{figure}[!b]
\centering
\includegraphics[width=0.9\linewidth]{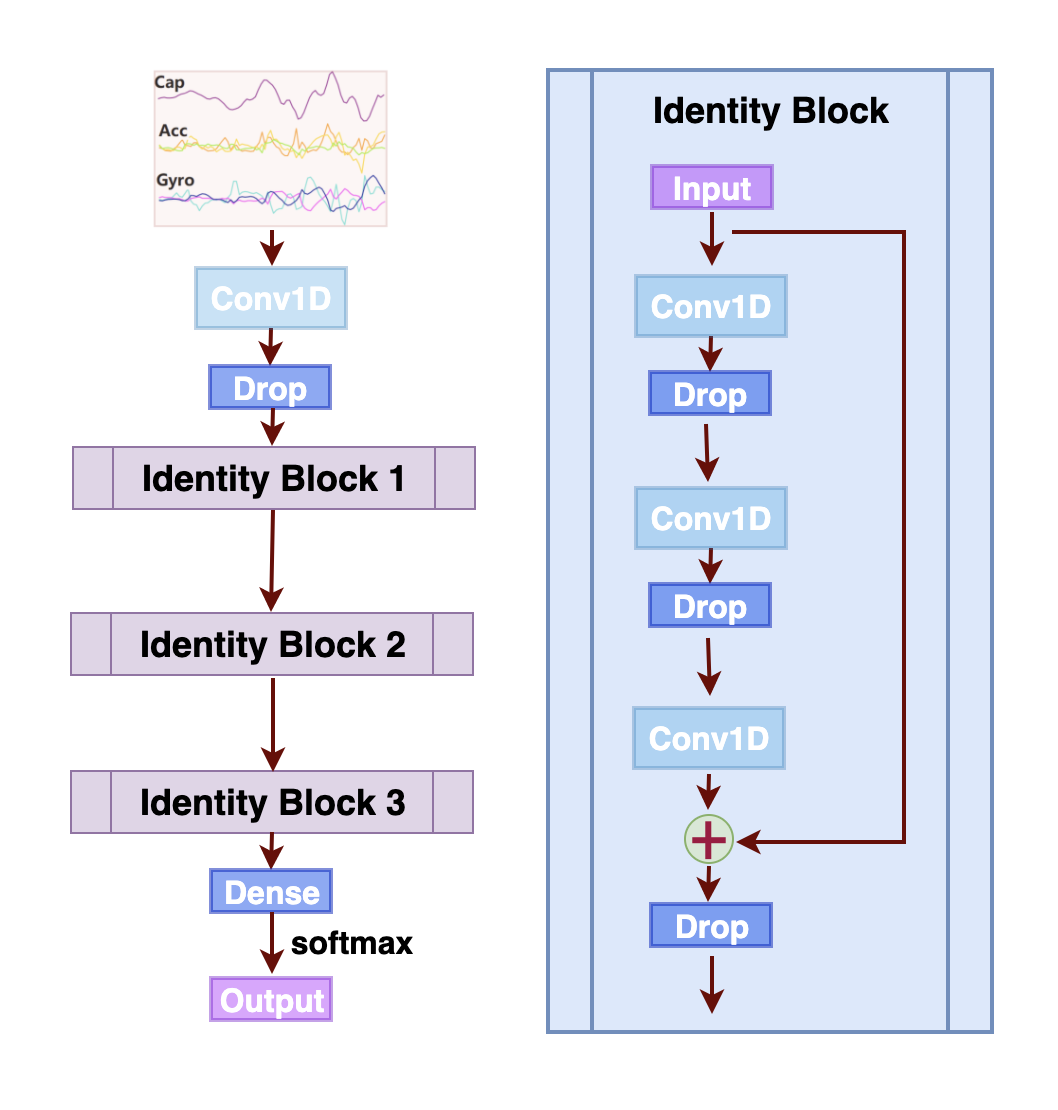}
\caption{Residual Deep Model}
\label{Model}
\end{figure}

For classification, we used a residual network composed of three identity blocks, following a 1D convolutional layer and followed by a dense layer (as Fig. \ref{Model} depicts). Each identity block has three 1D convolutional layers. One dimensional CNN has been proved to be an effective way for deriving features from a fixed-length segment of the overall time series data set, where the location of the features in the segment is not so important. All convolutions in this architecture use \textit{same} padding, which is convenient for going deeper as in our case each instance consists of a one-dimensional vector with window size ($40$) values. The input of the model is a seven-channel 1D data stream (7 x 40), which is normalized from the raw motion sensor data (three axes accelerometer and gyroscope, human body capacitance). The output is the predicted probability of the twelve classes (eleven workouts and one "Null" class).

\begin{table*}[t]
\centering
\begin{threeparttable}
\caption{Inference Performance on GAP8 and Cortex-M4/M7}
\label{Performance}

\begin{tabular}{ p{3.0cm} p{1.4cm} p{1.4cm} p{0.8cm} p{1.4cm} p{1.4cm} p{0.8cm} p{1.4cm} p{1.4cm}}


\hline
Platform & \multicolumn{2}{c}{GAP8} & &  \multicolumn{2}{c}{Cortex-M4} & &   \multicolumn{2}{c}{Cortex-M7}  \\ 
\hline
MCU &  \multicolumn{2}{c}{1xCV32E40P+8×CV32E40P} & &  \multicolumn{2}{c}{1×Cortex M4}  & &  \multicolumn{2}{c}{1×Cortex M7}  \\

Framework & \multicolumn{2}{c}{GapFlow}   &&  \multicolumn{2}{c}{CUBE.AI TFLite}  & & \multicolumn{2}{c}{CUBE.AI TFLite}  \\ 

Input Size & \multicolumn{2}{c}{40x7 }& &  \multicolumn{2}{c}{40x7} & & \multicolumn{2}{c}{40x7} \\ 

 MAC & \multicolumn{2}{c}{3,051,812}  & &  \multicolumn{2}{c}{3,039,408} & &  \multicolumn{2}{c}{3,039,408} \\ 
Flash &  \multicolumn{2}{c}{108.68kB}  &  &  \multicolumn{2}{c}{105.11kB}  & &  \multicolumn{2}{c}{105.11kB} \\ 
RAM & \multicolumn{2}{c}{33.37kB}   & & \multicolumn{2}{c}{14.41kB }  & &  \multicolumn{2}{c}{14.41kB}\\ 
Accuracy(quantized)  & \multicolumn{2}{c}{88.1}   & &  \multicolumn{2}{
c}{89.3}  &  &  \multicolumn{2}{c}{89.3} \\
Accuracy(Full Precision)  & \multicolumn{2}{c}{90.4}   & &  \multicolumn{2}{c}{90.4}  &  &  \multicolumn{2}{c}{90.4} \\

\hline
Clock & 80Mhz & Max(175Mhz) & &  60Mhz &  Max(120Mhz)  &  & 108 Mhz  &  Max(216Mhz)\\ 
\hline
Time/Inference[ms]  & 6.8 & 3.2 & &  114.25 & 60.36 & &  41.74 & 20.88 \\
Throughput [MMAC/s]  & 448.80 & 953.70 & &  26.60 & 50.35 & &  73.05 & 145.57\\
MAC/cycle  & 5.610 & 5.450 & & 0.44  & 0.42 & & 0.68  & 0.67\\
Power [mW]  & 54.60 & 129.36 &   & 47.16 & 85.67 & &  185.49 & 386.73 \\
Energy/inference [mJ]  & 0.37 & 0.41 &  & 5.39 & 5.17 &  & 7.74 & 8.07 \\
En. eff. [GMAC/s/W]  & 8.220 & 7.372 &  & 0.564 & 0.588 &  & 0.394  & 0.376 \\

\hline
\end{tabular}
\end{threeparttable}
\end{table*}

The model was trained using the categorical cross-entropy loss function and the Adam optimizer with $0.001$ learning rate and $0.9$ and $0.999$ for $\beta_1$ and $\beta_2$, respectively. Since the data set is highly imbalanced, containing more "Null" instances than any other exercise, every training instance is weighted based on the labels present inside the windows. The total weight of a window is inversely proportional to the frequency of its labels in the data set and is calculated based on the labels of a window.

The model is trained for $1000$ epochs with early stopping
using patience of $100$ to avoid overfitting.  To show that we can learn to recognize activities across subjects, we employed a leave-one-user-out procedure where, for each fold, the test set contains all days of one subject, while the training set contains all days for the remaining ones.
The TensorFlow Lite library was then used to convert the neural network model into the Tensorflow Lite file without quantization in this step. The optimization such as quantization will be performed by different frameworks specially designed for the embedded devices as described in the following subsection.

\section{Hardware Platforms}

For the evaluation of the inference performance on embedded processors, we chose three MCUs from STMicroelectronics and GreenWaves Technologies: STM32 Nucleo-144 development boards with Cortex-M4 processor and Cortex-M7 processor and GAP8, a RISC-V-based PULP platform. 

The Corte-M4 platform is equipped with 2 Mbytes of flash memory, 640 Kbytes of SRAM and can operate with a frequency of up to 120 MHz. The high-performance series of Cortex-M7 is equipped with 2 Mbytes of flash memory, 512 Kbytes of SRAM, and can be operated at up to 216 MHz frequency. For the conversion from the Tensorflow Lite model to the executable files on processors, we used the CUBE.AI tool from STMicroelectronics. The tool quantizes the pre-trained machine learning models, including neural network and traditional machine learning models, to reduce memory footprint and inference time, and integrates the generated optimized library into the specific project. Users could use a graphical interface or command line interface for the automation.

The PULP platform has two compute domains: a fabric controller with one CV32E40P core and a cluster domain with 8 CV32E40P cores. The fabric controller is the central system controller and resembles a standard MCU, which can delegate compute-intensive tasks to the cluster. The fabric controller has 512 kB L2 memory extendable via a HyperBus interface. The cluster is a multi-core compute domain with a shared 80 kB directly accessible L1 memory architecture and hardware thread synchronization, and enables highly efficient, parallel implementation of algorithms giving almost optimal linear speedup. Depending on the supplied voltage level, the frequency can be boosted to 250 MHz for the fabric controller and 175 Mhz for the cluster. We used the GAPflow framework for the conversion of the Tensorflow lite file. The framework comprises two core components: the NNTool, which parses the flatbuffer file (tensorflow Lite model) into an internal network description with topology optimization, like folding the batch normalization in the convolutional weights, and performs the quantization when requested; And the AutoTiler, which computes the optimal tiling scheme and converts the parsed model into optimized, debugable C code that can be executed efficiently in parallel by the eight cores in the cluster. To increase the efficiency of the memory usage, the Autotiler also utilizes fused layers generated from the NNTool, for example, a convolution layer followed by a pooling layer and the ReLU activation function can be performed by a single layer ConvPoolRelu supported by the AutoTiler.

\section{Experimental results}

Table \ref{Performance} presents the performance of the residual CNN model on the three hardware platforms with 8-bits quantization. Power consumption was measured by a power analyzer. The MAC operations of the proposed model are over three million and are slightly different on the platforms because of the various optimization strategy of each framework. The optimized model takes more than one hundred kilobytes of Flash space. Because of the parallel computing of the RISC-V cluster cores, the RAM space needed on GAP8 is larger than the Cortex-M platforms but still far behind the limitation (512kBytes). With the full precision inference, the workouts could be recognized with a balanced accuracy of 90.4\% on unregistered users. With a quantized model, the accuracy shows a minimal drop (88.1\% on GAP8 and 89.4\% on ARM Cortex M4/M7). The drop on the GAP8 platform could be negligible when looking at the achieved inference performance in time and power consumption on it, as the inference speed and energy consumed on GAP8 show significant improvement over the Cortex cores. Benefiting from the 8 RISC-V CV32E40P cluster, each inference takes 6.8ms with an 80Mhz clock and 3.2ms with the maximum clock on GAP8, which is 18.9x and 6.5x faster than the Cortex-M4 and Cortex-M7 while using the full clock speed, showing the feasibility of real-time on-board workouts recognition on the described data set with 20Hz sampling rate. The fast inference could also be indicated by the MAC operations in each cycle, where GAP8 offers 5.45 MAC operations per cycle, which shows a 13.0x and 8.1x improvement over the other two platforms. The energy consumed for each inference on GAP8 is 0.41mJ while it is 5.17mJ on Cortex-M4 and 8.07mJ on Cortex-M7 with the maximum clock speed. We also changed the clock on each platform for classification by using half of the full clock speed. It shows that the inference time and power consumption increase almost linearly as the clock increases. Still, the energy efficiency stays nearly the same on each platform since the time used for each inference also linearly decreases.

\section{Conclusion}

This work first presented a gym workouts data set for sensor-based activity recognition. The exploration of the automatic workouts recognition on three resource-constrained devices with a residual neural network was then described: two 32-bit microcontrollers with ARM-Cortex M4 and M7 core from STMicroelectronics and a RISC-V based open-source multi-core computing platform PULP from GreenWaves Technologies. Benefiting from the 8 RISC-V CV32E40P cluster, the inference performance on GAP8 shows significant advantages regarding the inference time, throughput, and energy efficiency over Cortex M4 and M7 cores while keeping the accuracy with minimal loss after quantization. With the maximal clock speed enabled, the GAP8 recognized the workouts with an accuracy of 88.1\% with a drop of 2.3\%, which is negligible considering the impressive performance improvement when compared with Cortex-M7 in inference time (3.2ms, 6.5x), throughput (953.70 MMAC/s, 6.6x), and energy efficiency(7.372 GMAC/s/W, 19.6x). This work demonstrated the feasibility of automatic gym workouts recognition directly on smart devices with limited hardware resources and battery life, and presented the outstanding inference performance on the open-source PULP platform than the classical Cortex-M platforms.


\bibliography{sample-base}
\bibliographystyle{IEEEtran}

\end{document}